\documentclass[aps,amsfonts,amsmath,prd,preprint,nofootinbib,tightenlines,showpacs]{revtex4}
\usepackage{epsf}

\def\be{\begin{equation}}
\def\ee{\end{equation}}
\def\bea{\begin{eqnarray}}
\def\eea{\end{eqnarray}}
\def\bml{\begin{subequations}}
\def\blea{\bml\begin{eqnarray}}
\def\elea{\end{eqnarray}\end{subequations}}

\begin{document}

\title{Energy conditions outside a dielectric ball}

\author{Noah Graham}
\email{ngraham@middlebury.edu} \affiliation{Department of Physics,
Middlebury College, Middlebury, VT  05753}
\author{Ken D.\ Olum}
\email{kdo@cosmos.phy.tufts.edu}
\author{Delia Schwartz-Perlov}
\email{Delia.Perlov@tufts.edu} \affiliation{Institute of
Cosmology, Department of Physics and Astronomy, Tufts University,
Medford, MA  02155}

\begin{abstract}
We show analytically that the vacuum electromagnetic stress-energy
tensor outside a ball with constant dielectric constant and
permeability always obeys the weak, null, dominant, and strong energy
conditions. There are still no known examples in quantum field theory
in which the averaged null energy condition in flat spacetime is
violated.
\end{abstract}


\maketitle

\section{Introduction}

A longstanding puzzle of general relativity is whether or not it is
possible to forbid the presence of exotic phenomena, such as the
construction of closed timelike curves \cite{cpc} and superluminal
travel \cite{Olum:1998mu}.  A variety of energy conditions exist that
would prevent these possibilities. The weak energy condition (WEC) and
null energy condition (NEC) respectively require that $T_{\mu\nu}
V^\mu V^\nu\ge0$ for any timelike and null vector $V^\mu$, but these
conditions have well-known violations even in free field theory in
Minkowski space.

The conditions may be weakened by requiring that they hold only
when averaged over a complete geodesic, and those averaged energy
conditions are still sufficient to rule out exotic phenomena.  The
averaged weak energy condition (AWEC) is violated in the standard
electromagnetic Casimir energy between parallel, perfectly
conducting planes, and it appears that the violation occurs even
with more realistic materials \cite{Sopova:2002cs}.  The system
does not violate the averaged null energy condition (ANEC),
however, because geodesics in directions which violate NEC all
pass through the conducting planes. Assuming any realistic
material conductor in these places, the positive stress-energy
there more than cancels the negative energy between the planes. In
\cite{Graham:2002} it was shown that for a simple model of scalar
fields, a static domain wall violates WEC, NEC and AWEC, even when
the domain wall energy is included, but again ANEC is obeyed, for
the same reason as above.

ANEC is violated in curved space-time
\cite{Viss96a,Viss96b,Viss96c,Viss97a}, but it is not clear that
such violations can be used to produce exotic phenomenon.  They
are typically overwhelmed by ordinary gravitational energy
densities, for objects that are not at the Planck scale.  No
violation of ANEC is known in a consistent quantum field theory in
flat spacetime.

One can try to avoid the problem of passing through the material in
the examples above by considering a geodesic passing outside a ball,
where one can safely ignore the contribution to the energy density
from the material of the ball.  For the case of minimally coupled
scalar fields, outside a perfectly reflecting spherical shell, one
finds the same results as in the domain wall case: WEC and NEC are
violated, but ANEC is obeyed \cite{Perlov:2003}.

In this paper, we analyze Casimir energy densities in the
electromagnetic field outside a ball with fixed dielectric
constant and permeability. The results also extend to the case of
a perfectly conducting sphere.\footnote{In the case of finite,
nonzero conductivity, we have absorption, coupling the incoming
waves to modes within the material.  The effects of this
interaction might depend in detail on the internal dynamics of the
material, so we do not consider it here.} In contrast to the
scalar field case, all energy conditions, including NEC (and
therefore ANEC) are obeyed everywhere outside the ball. Thus there
is still no counterexample to the conjecture that all quantum
field theories in flat spacetime obey ANEC.

The generality of our result --- depending only on the
requirement that the index of refraction $n$ is bigger than 1, which
is necessary to maintain causality in classical electromagnetism
without absorption --- suggests that it might be possible to extend it
to other cases in QED or other conformal theories.\footnote{In
the case of a conformal scalar field, using the techniques of
\cite{Perlov:2003} one finds that outside a perfectly reflecting
spherical boundary, NEC is also always obeyed. Unlike the
electromagnetic case, however, this result does not hold in each channel
individually; for the $\ell \ne 0$ partial wave contributions the NEC
is not always obeyed, but when one includes all $\ell$ it is obeyed
everywhere.}.

A number of related calculations have been carried out in other contexts.
The total energy of a dielectric ball has been studied, in particular with
applications to the controversial question of its relevance to
sonoluminescence \cite{Milton97,Nesterenko98}. However, it remains
difficult to formulate precisely because the energy is sensitive to the
details of the dynamics within the material.  These potential divergences
were studied in detail in \cite{Bordag98};  see also reviews
\cite{Bordag01,Milton99,Milton01}, and references therein.   The case of
the Casimir energy density outside a conducting sphere is a standard
calculation, originating with the work of Boyer \cite{Boyer68} and reviewed
in \cite{Mostepanenko97}.  Milton \cite{Milton80} calculates the total
energy including the interior and the net force on a dielectric sphere, and
in \cite{Milton04a,Milton04b} uses a comparison of the the stress density
inside and outside the sphere to find a Casimir contribution to the total
stress on the sphere.

In the next section, we give the energy density and the radial and
tangential pressures in terms of the electromagnetic field components.
In Section III we decompose the fields in vector spherical
harmonics, and in Section IV we derive the boundary conditions for the
ball with constant dielectric constant and permeability.  In section
V we quantize our fields, and in Section VI we compute the energy
density and pressures in the vacuum outside the ball.  In Section VII
we prove that all the energy conditions are obeyed, and we conclude in
Section VIII.

\section{The electromagnetic stress-energy tensor}

For the electromagnetic field outside a ball of radius $R$, with
constant dielectric constant and permeability, we will calculate the
vacuum expectation value of the energy density, radial and tangential
pressures. We will use Heaviside-Lorentz units throughout.

The symmetric electromagnetic stress-energy tensor is
\be
T^{\mu\nu}=g^{\mu\alpha}F_{\alpha\lambda}F^{\lambda\nu}
+\frac{1}{4}g^{\mu\nu}F_{\alpha\lambda}F^{\alpha\lambda}
\ee
The $00$ component is the electromagnetic energy density,
\be
T^{00} =\frac{1}{2}(E^2+B^2)
\ee
and the stresses are
\be
T^{ij} = -(E^i E^j+B^i B^j-\frac{1}{2}\delta^{ij}(E^2+B^2))
\ee
where $i,j$ are Cartesian coordinates in three space.  We can rewrite
the diagonal components in terms of a spherical coordinate system,
which we will decompose into radial and tangential components, denoted by
subscripts $r$ and $t$ respectively.  Let
\bea
\label{eqn:ffunctions} F_r &=& \frac{1}{2}(E_r^2+B_r^2)
\\ \label{eqn:ftunctions} F_t &= & \frac{1}{2} (E_t^2+B_t^2) \,.
\eea
Then the energy density and the radial and tangential pressures are
respectively given by
\bml
\label{eqn:gencomponents}
\bea
\rho = T^{00} &=& F_t+F_r\\
p_r = T^{rr} &=& F_t-F_r\\
p_t = T^{\phi\phi} &=& F_r \,.
\elea

\section{Classical fields}

We will work with the electromagnetic potential
$A^\mu(\mathbf{x},t)$.  In empty space we can work in the
radiation gauge, where $ \partial_\mu A^\mu = 0$ and $A^0=0$, so
that
\be
\label{eqn:EandA} \mathbf{E} = -\frac{\partial
\mathbf{A}}{\partial t}
\ee
and
\be
\label{eqn:BandA} \mathbf{B} =
\nabla \times \mathbf{A}\,.
\ee
We separate the two polarization
channels, so that we can write the vector potential as a sum of
classical modes,
\be
\label{eqn:vectorclassical} \mathbf{A}({\bf
x},t)= \sum_{\ell,m} \int_0^\infty \frac{dk}{\sqrt{2\omega}}
\left( \mathbf{A}_{k\ell m}^+ e^{-i\omega t} a_{k\ell m}^+ +
\mathbf{A}_{k\ell m}^- e^{-i\omega t} a_{k\ell m}^- +\text{complex
conjugate}\right)
\ee
where $\omega = k$ outside the ball since
the field is massless, the sum over $\ell$ and $m$ gives the
partial wave expansion in the $3$ spatial dimensions, and we
define
\bea
\mathbf{A}_{k\ell m}^+(r,\Omega) &\equiv& \mathbf{X}_{\ell m}(\Omega)
\psi^+_{k\ell} (r)\\
\mathbf{A}_{k\ell m}^-(r,\Omega) &\equiv& \frac{1}{k}\nabla \times
\left(\mathbf{X}_{\ell m}(\Omega) \psi^-_{k\ell}(r)\right)
\eea
in terms of the vector spherical harmonics
\be
\mathbf{X}_{\ell m}(\Omega)\equiv \frac{-i\left(\mathbf{r}\times \nabla \right)
Y_{\ell m}(\Omega)}{\sqrt{\ell(\ell+1)}} \,.
\ee
The modes
$\mathbf{A}^+_{k \ell m}$ are the magnetic multipoles, also known
as the transverse electric (TE) modes, and $ \mathbf{A}^-_{k \ell
m}$ are the electric multipoles, or transverse magnetic (TM)
modes. The radial functions $\psi^+_{k\ell}(r)$ and
$\psi^-_{k\ell}(r)$ satisfy the same equation of motion as a
scalar field,
\be
\label{eqn:Schrodinger} \left(-\frac{d^2}{dr^2} -
\frac{2}{r}\frac{d}{dr} + \frac{\ell(\ell+1)}{r^{2}}\right)
\psi^\pm_{k\ell}(r) = k^2 \psi^\pm_{k\ell}(r)\,.
\ee
For any function $g_{k\ell}(r)$ obeying this equation, we have the
following useful identities:
\bea
\mathbf{\hat r} \cdot \mathbf{X}_{\ell m} g_{k\ell}(r) &=& 0 \\
\mathbf{\hat r} \cdot \left(\nabla\times
\left(\mathbf{X}_{\ell m} g_{k\ell}(r) \right) \right)
&=&\frac{i}{r}\sqrt{\ell(\ell+1)} g_{k\ell}(r)Y_{\ell m} \\
\label{eqn:identity3} \nabla \times\left(\nabla \times
\left(\mathbf{X}_{\ell m} g_{k\ell}(r)\right) \right)
&=& -\nabla^{2} \left(\mathbf{X}_{\ell m} g_{k\ell}(r)\right)
= k^2 \mathbf{X}_{\ell m} g_{k\ell}(r)\\
\mathbf{\hat r} \times\left(\nabla \times \left(\mathbf{X}_{\ell
m} g_{k\ell}(r)\right)\right) &=&
-\frac{1}{r}\frac{\partial}{\partial r} (r
g_{k\ell}(r))\mathbf{X}_{\ell m}\,.
\eea

The $\mathbf{E}$ and $\mathbf{B}$ fields, which are related to
$\mathbf{A}$ via Eqs.\ (\ref{eqn:EandA}) and (\ref{eqn:BandA})
can then be expressed
\be
\mathbf{E}({\bf
x},t)= \sum_{\ell,m} \int_0^\infty \frac{dk}{\sqrt{2\omega}}
\left( \mathbf{E}_{k\ell m}^+ e^{-i\omega t} a_{k\ell m}^+ +
\mathbf{E}_{k\ell m}^- e^{-i\omega t} a_{k\ell m}^-
+\text{complex conjugate}\right)
\ee
and likewise for $\mathbf{B}$, with
\bea
\mathbf{E}_{k\ell m}^+ &=& i\omega\mathbf{X}_{\ell m} \psi^+_{k\ell}(r)\\
\mathbf{B}_{k\ell m}^+ &=& \nabla \times \left(\mathbf{X}_{\ell
m} \psi^+_{k\ell}(r)\right)\\
\mathbf{E}_{k\ell m}^- &=& i\omega \frac{1}{k}
\nabla \times \left(\mathbf{X}_{\ell m} \psi^-_{k\ell}(r)\right) \\
\mathbf{B}_{k\ell m}^- &=& \nabla \times
\left(\frac{1}{k} \nabla \times
\left(\mathbf{X}_{\ell m} \psi^-_{k\ell}(r)\right) \right)
= k \mathbf{X}_{\ell m} \psi^-_{k\ell}(r) \,.
\eea

Using the identities above, we can now decompose the modes separately
into radial and tangential components,
\bea
E_{k\ell m, r}^+ &=& 0\\
\mathbf{E}_{k\ell m, t}^+ &=& i\omega\mathbf{X}_{\ell m} \psi^+_{k\ell}(r)\\
B_{k\ell m, r}^+ &=&
\frac{i}{r} \sqrt{\ell(\ell+1)}Y_{\ell m} \psi^+_{k\ell}(r)\\
\mathbf{B}_{k\ell m, t}^+ &=& \frac{1}{r}\frac{\partial}{\partial
r}\left(r\psi^+_{k\ell}(r)\right)\mathbf{\hat r}\times
\mathbf{X}_{\ell m}
\eea
and
\bea
E_{k\ell m,r}^-&=& -\frac{\omega}{kr}\sqrt{\ell(\ell+1)}Y_{\ell
m} \psi^-_{k\ell}(r)\\
\mathbf{E}_{k\ell m,t}^- &=&  \frac{i\omega}{kr}
\frac{\partial}{\partial r}\left(r\psi^-_{k\ell}(r)\right)
\mathbf{\hat r}\times \mathbf{X}_{\ell m}\\
B_{k\ell m,r}^- &=& 0\\
\mathbf{B}_{k\ell m,t}^- &=&
k  \mathbf{X}_{\ell m} \psi^-_{k\ell}(r)  \,.
\eea

\section{Boundary conditions for the uniform ball}

The general solution to the radial equation of motion, Eq.\
(\ref{eqn:Schrodinger}), can be written
\be
\label{eqn:wavefn} \psi^\pm_{k\ell}(r)
=\frac{k}{\sqrt{2\pi}}\left[e^{2i\delta^\pm_\ell (k)}
h^{(1)}_\ell(z)+h^{(2)}_\ell(z)\right]
\ee
where $z = kr$, $\delta^{\pm}_\ell(k)$ is the scattering phase shift,
and
\be
h^{(1,2)}_\ell(z) =\sqrt{\frac{\pi}{2z}} H^{(1,2)}_{\ell+1/2}(z)
\ee
are the spherical Hankel functions.

We now consider a uniform ball of radius $R$, dielectric constant
$\epsilon$, permeability $\mu$, and zero conductivity. Nonzero and
finite conductivity leads to absorption, in which case the Casimir
energy may depend in detail on the properties of the conducting
material, so we will not consider that possibility. The case of
perfect conductivity, however, can be obtained as the limit where
$\epsilon \to \infty$.

Within the ball, we have the same decomposition of the fields,
except there we have $k = n \omega$, where $n=\sqrt{\epsilon\mu}$
is the index of refraction.  We will consider only cases with
index of refraction $n\ge1$, since a frequency-independent index
of refraction below 1 would violate causality.\footnote{A real
material can have $n<1$ at some frequencies, but such materials
necessarily have absorption as well.}  We can also scale the
wavefunction inside the ball by an overall constant in each
channel.  At the surface, $\epsilon E_r$, $\mathbf{E}_t$, $B_r$,
and $\frac{1}{\mu} \mathbf{B}_t$ are continuous, giving two
independent conditions for the two unknown quantities: the phase
shift and the normalization constant of the interior wavefunction.
Solving for the TE phase shift, we have \cite{Newton66}
\be
e^{2i\delta^+_{\ell}(k)}= -\frac{\alpha \widehat
\jmath_\ell{}'(nx)\widehat h^{(2)}_\ell(x) -\widehat
\jmath_\ell(nx)\widehat h^{(2)}_\ell{}'(x)} {\alpha \widehat
\jmath_\ell{}'(nx)\widehat h^{(1)}_\ell(x) -\widehat
\jmath_\ell(nx)\widehat h^{(1)}_\ell{}'(x)}
\ee
where prime denotes differentiation with respect to the function's
argument, and we have defined $\alpha =\sqrt{\epsilon/\mu}$, $x=kR$,
and the Riccati-Bessel and Riccati-Hankel functions
\bea
\widehat \jmath_\ell(z)&=&z j_\ell(z)=\sqrt{\frac{\pi z}{2}}
J_{\ell+1/2} (z)\\
\widehat h^{(1,2)}_\ell(z)&=&z h^{(1,2)}_\ell(z) =
\sqrt{\frac{\pi z}{2}} H^{(1,2)}_{\ell+1/2}(z) \,.
\eea

Defining
\be
s(x) = \frac{\widehat
\jmath_\ell{}'(x)}{\widehat \jmath_\ell(x)}
\ee
we can write
\be
\label{eqn:smat} e^{2i\delta^+_{\ell}(k)}= -\frac{\alpha
s(nx)\widehat h^{(2)}_\ell(x) -\widehat h^{(2)}_\ell{}'(x)} {\alpha
s(nx)\widehat h^{(1)}_\ell(x) -{\widehat h^{(1)}_\ell}{}'(x)}
\ee
and thus
\be
\label{eqn:S1} e^{2i\delta^+_{\ell}(k)}-1=
-2\frac{\alpha s(nx)\widehat \jmath_\ell(x) -\widehat
\jmath_\ell{}'(x)} {\alpha s(nx)\widehat h^{(1)}_\ell(x)
-\widehat h^{(1)}_\ell{}'(x)}
\ee
The TM mode is the same with
$\alpha\to1/\alpha$, and $\delta^+_{\ell}(k) \to
\delta^-_{\ell}(k)$.

\section{Quantization}

To quantize the electromagnetic field we simply declare that the
coefficients $a_{k\ell m}^\pm$ in Eq.\ (\ref{eqn:vectorclassical})
are operators and take the Hermitian conjugate of those operators
in the complex conjugate term.  For the operators to have the
usual relations in which all commutators vanish except \be
[a_{k\ell m}, a_{k'\ell'm'}^\dagger] = \delta(k-k') \delta_{\ell
\ell'} \delta_{m m'} \,. \ee We require the radial wave functions
be normalized by \be \label{eqn:norm} \int_0^\infty r^{2}
\psi^\pm_{k\ell}(r)^* \psi^\pm_{k'\ell}(r) \, dr = \delta(k-k')
\ee which is satisfied by Eq.\ (\ref{eqn:wavefn}).  The vector and
scalar spherical harmonics are normalized by the completeness
relations \bea \sum_{m} \left|\mathbf{X}_{\ell m}\right|^2 =
\sum_{m} \left|Y_{\ell m}\right|^2 = \frac{2\ell+1}{4\pi} \,. \eea
The quantum field $\mathbf{A}(\mathbf{x},t)$ then obeys the
conventional equal-time commutation relations in radiation gauge,
\be \left[A_i(\mathbf{x},t), E_j(\mathbf{y},t)\right] = -i\left(
\delta_{ij} - \frac{\partial_i \partial_j}{\nabla^2} \right)
\delta(\mathbf{x} - \mathbf{y}) \,. \ee Since we are interested in
computing the vacuum expectation values of the components of the
electromagnetic field, from this point on, notations such as
$E_r^2$ and $F_r$ will denote the quantum mechanical vacuum
expectation values of those quantities.

For the TE mode we find
\bea
E_{+r}^2 & = & 0   \\
E_{+t}^2 & = &  \sum_{\ell=1}^\infty
\frac{2\ell+1}{8\pi} \int_0^\infty dk\,\omega
\left| \psi^+_{k\ell}(r)\right|^2\\
B_{+r}^2 & = & \sum_{\ell=1}^\infty
\frac{2\ell+1}{8\pi} \frac{\ell(\ell+1)}{r^2}
\int_0^\infty\frac{dk}{\omega} \left| \psi^+_{k\ell}(r)\right|^2 \\
B_{+t}^2
& = & \sum_{\ell=1}^\infty \frac{2\ell+1}{8\pi} \int_0^\infty
\frac{dk}{\omega} \left|\frac{1}{r}
\frac{\partial}{\partial r} \left(r
\psi^+_{k\ell}(r)\right)\right|^2\,.
\eea

The renormalized contribution is found by subtracting the free
wave contribution, $\psi^{+(0)}_{k\ell}(r)$, yielding
\bea
E_{+r}^2 & = & 0   \\
E_{+t}^2 & = &  \sum_{\ell=1}^\infty
\frac{2\ell+1}{8\pi} \int_0^\infty dk\,\omega
\left(\left| \psi^+_{k\ell}(r)\right|^2-\left|
\psi^{+(0)}_{k\ell}(r)\right|^2\right)\\
B_{+r}^2 & = & \sum_{\ell=1}^\infty \frac{2\ell+1}{8\pi}
\frac{\ell(\ell+1)}{r^2} \int_0^\infty\frac{dk}{\omega} \left(
\left| \psi^+_{k\ell}(r)\right|^2-\left|
\psi^{+(0)}_{k\ell}(r)\right|^2\right)\\
B_{+t}^2
& = & \sum_{\ell=1}^\infty \frac{2\ell+1}{8\pi} \int_0^\infty \frac{dk}{\omega}
 \left(\left|\frac{1}{r}\frac{\partial}{\partial r} \left(r
\psi^+_{k\ell}(r)\right)\right|^2 -
\left|\frac{1}{r} \frac{\partial}{\partial r}
\left(r \psi^{+(0)}_{k\ell}(r)\right)\right|^2\right)
\eea

Using the general form of the wavefunction, Eq.\ (\ref{eqn:wavefn}),
and subtracting the free wavefunctions given by Eq.\
(\ref{eqn:wavefn}) with $\delta_\ell=0$, we find
\be
\label{eqn:diffwavefn}
|\psi^\pm_{k\ell}(r)|^2-|\psi^{\pm(0)}_{k\ell}(r)|^2=
\frac{k^2}{2\pi}\left[ \left(e^{2i
\delta^\pm_\ell}-1\right)h^{(1)}_\ell(z)^2 +\left(e^{-2i
\delta^\pm_\ell}-1\right)h^{(2)}_\ell(z)^2\right]\,.
\ee
and
\be\label{eqn:diffwavefnderiv}
\left|\frac{1}{r}\frac{\partial}{\partial r}
\left(r\psi^\pm_{k\ell}(r)\right)\right|^2
-\left|\frac{1}{r}\frac{\partial}{\partial r}
\left(r \psi^{\pm(0)}_{k\ell}(r)\right)\right|^2 =\frac{k^2}{2\pi r^2}
\left[\left(e^{2i\delta^\pm_\ell}-1\right) \widehat h^{(1)}_\ell{}'(z)^2
 + \left(e^{-2i\delta^\pm_\ell}-1\right)\widehat h^{(2)}_\ell{}'(z)^2\right]
\ee
 where prime denotes differentiation with respect to $z=kr$,
and we write $\delta_\ell$ instead of $\delta_\ell(k)$ for
simplicity. The Appendix of \cite{Perlov:2003} showed that the
second term of Eq.\ (\ref{eqn:diffwavefn}) is just the first term
with the replacement $k\rightarrow-k+i\epsilon$ and likewise for
Eq.\ (\ref{eqn:diffwavefnderiv}).  Thus we can drop the second
terms in both equations and extend the range of integration over
$k$ to $-\infty$, with the understanding that $k$ is to be taken
above any branch cut on the negative real axis,
\bml\label{eqn:tecomponents}
\bea
E_{+r}^2 & = & 0   \\
E_{+t}^2 & = &  \sum_{\ell=1}^\infty \frac{2\ell+1}{16\pi^2}
\int_{-\infty}^\infty dk\, \omega k^2
\left(e^{2i \delta_{\ell}^+}-1\right)h^{(1)}_\ell(z)^2\\
B_{+r}^2 & = & \sum_{\ell=1}^\infty \frac{2\ell+1}{16\pi^2r^2}
 \ell(\ell+1) \int_{-\infty}^\infty \frac{dk}{\omega}
k^2 \left(e^{2i \delta_{\ell}^+}-1\right)h^{(1)}_\ell(z)^2 \\
\label{eqn:TEparBcomponent} B_{+t}^2 & = & \sum_{\ell=1}^\infty
\frac{2\ell+1}{16\pi^2 r^2} \int_{-\infty}^\infty
\frac{dk}{\omega} k^2\left(e^{2i \delta_{\ell}^+}-1\right)
\widehat h^{(1)}_\ell{}'(z)^2
\elea

The calculation for the TM modes is exactly analogous,
\bml\label{eqn:tmcomponents}
\bea
E_{-r}^2 & = & \sum_{\ell=1}^\infty
\frac{2\ell+1}{16\pi^2 r^2} \ell(\ell+1)
\int_{-\infty}^\infty \frac{dk}{\omega}
k^2\left(e^{2i \delta_{\ell}^-}-1\right)h^{(1)}_\ell(z)^2  \\
E_{-t}^2 & = & \sum_{\ell=1}^\infty \frac{2\ell+1}{16\pi^2 r^2}
\int_{-\infty}^\infty \frac{dk}{\omega} k^2\left(e^{2i
\delta_{\ell}^-}-1\right)
\widehat h^{(1)}_\ell{}'(z)^2 \\
B_{-r}^2 & = & 0   \\
B_{-t}^2 & = &  \sum_{\ell=1}^\infty \frac{2\ell+1}{16\pi^2}
\int_{-\infty}^\infty dk\, \omega k^2 \left(e^{2i
\delta_{\ell}^-}-1\right)h^{(1)}_\ell(z)^2
\elea

\section{Energy density and pressure}
Substituting Eqs.\ (\ref{eqn:tecomponents}) and
(\ref{eqn:tmcomponents}) into Eqs. (\ref{eqn:ffunctions}) and
(\ref{eqn:ftunctions}) we find
\bea
F_r &=& \frac{1}{32\pi^2
r^2}\sum_{\ell=1}^\infty (2\ell+1)\ell(\ell+1)
\int_{-\infty}^\infty dk\, T_k
\frac{k^2}{\omega} h^{(1)}_\ell(z)^2\\
F_t &=& \frac{1}{32\pi^2 r^2}\sum_{\ell=1}^\infty (2\ell+1)
\int_{-\infty}^\infty dk\, T_k \left\{\omega \widehat
h^{(1)}_\ell(z)^2+\frac{k^2}{\omega} \widehat
h^{(1)}_\ell{}'(z)^2\right\}
\eea
where
\be
T_k = \left(e^{2i\delta_{\ell}^+(k)}-1\right) +
\left(e^{2i\delta_{\ell}^-(k)}-1\right)
\ee
Together, these expressions reproduce the energy density found in
\cite{Mostepanenko97}.

Following the methods used in \cite{Graham:2002} and
\cite{Perlov:2003}, we now convert this expression to a contour
integral, which we close in the upper half plane.  Since our
interaction has finite range and no bound states, the $S$-matrix
in Eq.\ (\ref{eqn:smat}) is analytic everywhere in the upper
half-plane \cite{Newton66}.  Thus the only contribution to the
integral comes from the branch cut along the imaginary $k$ axis
coming from $\omega = \sqrt{k^2 + i\epsilon}\,$.   To the right
$\omega = \sqrt{k^2} = k$, but to the left $\omega = -k$, so with
$k = i\kappa$, we obtain
\bea
\label{eqn:Fr} F_r &=&-\frac{1}{4\pi^3 r^2}\sum_{\ell=1}^\infty
(2\ell+1)\ell(\ell+1) \int_{0}^\infty d\kappa\,
\tilde{T}_\kappa  \kappa  k_\ell(\zeta)^2 \\
\label{eqn:Ftim} F_t &=& \frac{1}{4\pi^3r^2}\sum_{\ell=1}^\infty (2\ell+1)
\int_{0}^\infty d\kappa\, \tilde{T}_\kappa \kappa \left\{\widehat
k_\ell(\zeta )^2 - \widehat k_\ell{}'(\zeta)^2\right\}
\eea
where $\zeta =\kappa r$,
\be
\label{eqn:RBk} \widehat k_\ell(\zeta)=\zeta
k_\ell(\zeta) = \sqrt{\frac{\pi \zeta}{2}} K_{\ell+1/2} (\zeta)
\ee
is the modified Riccati-Bessel function of the third kind, and
\be
\tilde T_\kappa = \frac{(-)^\ell}{\pi} \left[
\left(e^{2i\delta_{\ell}^+(i\kappa)}-1\right) +
\left(e^{2i\delta_{\ell}^-(i\kappa)}-1\right) \right] \,.
\ee
From Eq.\ (\ref{eqn:S1}), we get
\be
\tilde T_\kappa =\frac{\alpha f(n\chi)\widehat \imath_\ell(\chi) -\widehat
\imath_\ell{}'(\chi)} {\alpha f(n\chi)\widehat k_\ell(\chi)
-\widehat k_\ell{}'(\chi)} + \left\{ \alpha\to1/\alpha \right\}
\ee
where $\chi =\kappa R$,
\be
\label{eqn:f}
f (\chi) = s (i\chi)=
\frac{\widehat \imath_\ell{}' (\chi)}{\widehat \imath_\ell(\chi)}
\ee
and
\be
\label{eqn:RBi} \widehat \imath_\ell(\zeta)=\zeta
i_\ell(\zeta)=\sqrt{\frac{\pi \zeta}{2}} I_{\ell+1/2} (\zeta)
\ee
is the modified Riccati-Bessel function of the first kind.

\section{Positivity}

We would like to know the sign of $F_r$ and $F_t$.  First consider
$F_r$.  Everything on the right hand side of Eq.\ (\ref{eqn:Fr}) is
manifestly positive, except for $\tilde T_\kappa$ and the
overall negative sign, so $F_r$ has the opposite sign from $\tilde
T_\kappa$.

Now consider $F_t$.  The properties of modified Riccati-Bessel functions are discussed in
Appendix \ref{sec:RB}.  From Eq.\ (\ref{eqn:g1}) we find
\be
\widehat k_\ell{}'(\zeta) < -\widehat k_\ell(\zeta )
\ee
Since both sides are negative,
\be
\widehat k_\ell{}'(\zeta)^2 > \widehat k_\ell(\zeta )^2
\ee
so the term in braces in Eq.\ (\ref{eqn:Ftim}) is negative, and $F_t$
also has the opposite sign from $\tilde T_\kappa$.

Let us now determine the sign of $\tilde T_\kappa$.  We
can write $\tilde T_\kappa$ in the form
\be
\label{eqn:Tabcd}
\frac{\alpha a-b}{\alpha c-d}
+\frac{\alpha^{-1}a-b}{\alpha^{-1}c-d} =
\frac{2ac+2bd-\left(\alpha +\alpha^{-1}\right) (ad+bc)}
{c^2+d^2-\left(\alpha +\alpha^{-1}\right)cd} \ee with \bea
a &=& f(n\chi)\ \widehat \imath_\ell(\chi)\\
b &=& \widehat \imath_\ell{}'(\chi)\\
c &=& f(n\chi)\widehat k_\ell(\chi)\\
d &=& \widehat k_\ell{}'(\chi)
\eea

The functions $\widehat \imath_\ell$ and $\widehat k_\ell$ are
positive, and $\widehat \imath_\ell$ is increasing, so $f$ is
positive and $a$, $b$, and $c$ are positive.  On the other hand,
$\widehat k_\ell$ is a decreasing function, so $d$ is negative.
Thus each term in the denominator of Eq.\ (\ref{eqn:Tabcd})
contributes positively.

We can write the numerator of Eq.\ (\ref{eqn:Tabcd}) as
\be
\label{eqn:Tnum} 2(a-b)(c - d)-\left(\alpha
+\alpha^{-1}-2\right) (ad+bc) \ee and we have \be\label{eqn:adbc}
ad+bc=f(n\chi)\left[\widehat \imath_\ell(\chi) \widehat
k_\ell'(\chi) +\widehat k_\ell(\chi)\widehat
\imath_\ell'(\chi)\right]
\ee
By multiplying Eq.\ (\ref{eqn:fg}) by the positive quantity $\widehat
\imath_\ell(\chi) \widehat k_\ell(\chi)$ we find that the term in
brackets in Eq.\ (\ref{eqn:adbc}) is positive.  Since $f(n\chi) > 0$,
and $\left(\alpha +\alpha^{-1}-2\right)=(\alpha-1)^2/\alpha > 0$, the
second term in Eq.\ (\ref{eqn:Tnum}) contributes negatively.

Now $c - d$ is manifestly positive, and
\be
a-b = \widehat \imath_\ell(\chi)\left[ f(n\chi) - f (\chi)\right]
\ee
Since $n > 1$ by assumption, Eq.\ (\ref{eqn:fprime}) tells us that the
term in brackets is negative, so the first term in Eq.\
(\ref{eqn:Tnum}) is negative.  Thus $\tilde T_\kappa< 0$, and both
$F_r$ and $F_t$ are positive.

From Eqs.\ (\ref{eqn:gencomponents}), we see that $\rho > p_t >0$
and $\rho> |p_r|$.  Thus the weak, null, dominant, and strong
energy conditions are satisfied at each point, and therefore the
corresponding averaged energy conditions are satisfied on every
geodesic that lies outside the ball.

\section{Discussion}

We have demonstrated, without resorting to numerical calculations,
that the electromagnetic Casimir energy density is positive
outside a ball with constant dielectric constant and permeability,
and that all the standard energy conditions are obeyed. We needed
to assume only that the index of refraction is greater than 1,
which is necessary to maintain causality.  We did not consider the
cases of absorption and dispersion. We expect, however, that our
results would hold in such cases, and furthermore, we expect them
to remain valid in the case of material properties that vary with
radius.

The electromagnetic situation discussed here gives a simpler
result than the case of a scalar field.  For the minimally coupled
scalar field \cite{Perlov:2003}, the pointwise energy conditions
are not obeyed, although the average null energy condition
nevertheless holds for every geodesic outside the ball.  For the
conformally coupled case the null energy condition does hold at
each point, but only when all angular momentum modes are
considered.  In the electromagnetic case, the energy conditions
hold at each point separately for each $k$ and $l$.

\section{Acknowledgments}
We would like to thank Larry Ford and Robert L. Jaffe for helpful
conversations.  K.\ D.\ O. was supported in part by the National
Science Foundation (NSF).  N.\ G.\ was supported in part by the NSF
through the Vermont Experimental Program to Stimulate Competitive
Research (VT-EPSCoR).

\appendix

\section{Riccati-Bessel function inequalities}
\label{sec:RB}
We let $\widehat \imath_\ell$ and  $\widehat k_\ell$ be modified Riccati-Bessel
functions as in Eqs.\ (\ref{eqn:RBi}) and (\ref{eqn:RBk}), and
define their logarithmic derivatives,
\be
f (x) = \frac{\widehat \imath_\ell{}' (x)}{\widehat \imath_\ell(x)}
\ee
as in Eq.\ (\ref{eqn:f}) and
\be
g(x) = \frac{\widehat k_\ell' (x)}{\widehat k_\ell(x)}
\ee
Now the modified Riccati-Bessel functions satisfy the differential equation
\be
x^2{\cal F}''_\ell-(x^2+\ell (\ell+1)){\cal F}_\ell= 0
\ee
with ${\cal F}=\widehat \imath$ or $\widehat k$, so the derivatives satisfy
a generalized Riccati equation \cite{Watson96},
\be\label{eqn:hprime}
h'(x) =\frac{{\cal F}_\ell'' (x)}{{\cal F}_\ell(x)}-\frac{{\cal
F}_\ell' (x)^2}{{\cal F}_\ell(x)^2}
=\frac{\ell (\ell+1)}{x^2} +1-h(x)^2
\ee
for $h = f$ or $g$.

Differentiating Eq.\ (\ref{eqn:hprime}) gives
\be
h''(x) =- 2\frac{\ell (\ell+1)}{x^3}-2h(x)h'(x)
\ee
from which we conclude that if $h'(x) = 0$ for some $x$ then $h''(x) <
0$ there.  Thus $h'(x)$ cannot increase through 0, and so if $h'(x)$
is negative for some $x$, then it is negative for all larger $x$,
while if it is positive for some $x$, then it must be positive for all
smaller $x$.

Now for $x\to0$, $f(x)\to (l+1)/x$, and so for small $x$, $f'(x)$ is
negative, so
\be\label{eqn:fprime}
f'(x)< 0
\ee
for all $x$.  As $x\to\infty$, $f (x)\to1$, so
\be
f(x) > 1
\ee
for all $x$.

For $x\to\infty$,
\be
\widehat k_\ell(x)\to\frac{\pi}{2}
e^{-x}\left(1+\frac{\ell(\ell+1)}{2x} +O (x^{-2})\right) \ee so
\be \widehat k'_\ell(x)\to -\widehat k_\ell(x)-\frac{\pi}{2}
e^{-x}\left(\frac{\ell(\ell+1)}{2x^2} +O (x^{-3})\right)
\ee
therefore
\be
g(x)\to-1-\frac{\ell(\ell+1)}{2x^2}+O (x^{-3})
\ee
Since $g'(x)>0$ for large $x$, we have
\be
\label{eqn:gprime}
g'(x)>0
\ee
for all $x$, and thus
\be
\label{eqn:g1} g(x) < -1
\ee
for all $x$.

Now by subtracting Eq.\ (\ref{eqn:hprime}) for $f$ from the same
equation for $g$, we find
\be
g'(x)-f'(x)=f(x)^2-g(x)^2
\ee
From Eqs.\ (\ref{eqn:fprime}) and (\ref{eqn:gprime}) this quantity is
positive, so $f(x)^2 > g(x)^2$.  Since $g(x) < 0$ while $f(x)> 0$, we
find
\be\label{eqn:fg}
f(x)+g(x) > 0
\ee

\bibliographystyle{apsrev}
\bibliography{gr}

\end{document}